\documentclass[reprint,amsmath,amssymb,aps,pre,floatfix]{revtex4-1}
\usepackage{graphicx}
\usepackage{comment}
\usepackage{dcolumn}
\usepackage{bm}
\usepackage{amsmath}
\usepackage{qcircuit}
\usepackage{color}
\usepackage{xypic}
\begin{document}

\title{Quantum Classical Algorithm for the Study of Phase Transitions in the Hubbard Model via Dynamical Mean-Field Theory }

\author{Anshumitra Baul (\href{mailto:abaul1@lsu.edu}{abaul1@lsu.edu})} 
\affiliation{Department of Physics and Astronomy, Louisiana State University, Baton Rouge, Louisiana 70803, USA}

\author{Herbert F Fotso}
\affiliation{Department of Physics, University at Buffalo,  Buffalo, NY 14260, USA}

\author{Hanna Terletska}
\affiliation{Department of Physics and Astronomy, Middle Tennessee State University, Murfreesboro, TN 37132, USA}

\author{Juana Moreno}
\affiliation{Department of Physics and Astronomy, Louisiana State University, Baton Rouge, Louisiana 70803, USA}%
\affiliation{Center for Computation and Technology, Louisiana State University, Baton Rouge, LA 70803, USA}

\author{Ka-Ming Tam}
\affiliation{Department of Physics and Astronomy, Louisiana State University, Baton Rouge, Louisiana 70803, USA}%
\affiliation{Center for Computation and Technology, Louisiana State University, Baton Rouge, LA 70803, USA }

\date{\today}

\begin{abstract}

Modeling many-body quantum systems is widely regarded as one of the most promising applications for near-term noisy quantum computers. However, in the near term, system size limitation will remain a severe barrier for applications in materials science or strongly correlated systems. A promising avenue of research is to combine many-body physics with machine learning for the classification of distinct phases. We present a workflow that synergizes quantum computing, many-body theory, and quantum machine learning(QML) for studying strongly correlated systems. In particular, it can capture a putative quantum phase transition of the stereotypical strongly correlated system, the Hubbard model.
Following the recent proposal of the hybrid classical-quantum algorithm for the two-site dynamical mean-field theory(DMFT), we present a modification that allows the self-consistent solution of the single bath site DMFT. The modified algorithm can easily be generalized for multiple bath sites. 
This approach is used to generate a database of zero-temperature wavefunctions of the Hubbard model within the DMFT approximation. We then use a QML algorithm to distinguish between the metallic phase and the Mott insulator phase to capture the metal-to-Mott insulator phase transition. We train a recently proposed quantum convolutional neural network(QCNN) and then utilize the QCNN as a quantum classifier to capture the phase transition region. This work provides a recipe for application to other phase transitions in strongly correlated systems and represents an exciting application of small-scale quantum devices realizable with near-term technology.

\end{abstract}

\maketitle

\section{Introduction}

Strong electronic interactions in quantum materials give rise to many phenomena that potentially harbor properties suitable for technological applications. Some notable effects include the metal-insulator transition, heavy fermions, fractional quantum effects, frustrated magnetism, and non-Fermi liquid metals \cite{Varma_etal_2002, Savary_Balents_2016, Stewart_1984}. However, delving into the realm of strongly correlated systems poses significant challenges, primarily due to the absence of a suitable starting point for perturbative methods. This challenge is epitomized by the elusive nature of understanding the superconducting property of cuprates despite decades of research. \cite{Proust_Taillefer_2019, Varma_2020}. 
A plethora of classical numerical methods have been employed to solve simplified models of strongly correlated systems, among various other important models, such as the Hubbard model.\cite{Kanamori_1963,Hubbard_I_1963}. However, these numerical methods are often hindered by the minus sign problem for Quantum Monte Carlo (QMC), or by the exponential growth of the Hilbert space with the system sizes for exact diagonalization. This
prevents simulations from attaining low temperatures. Amid these challenges, embedding schemes like dynamical mean field theory (DMFT) \cite{georges1992hubbard,DMFT_RMP,Muller_Hartmann_1989a,Muller_Hartmann_1989b,Metzner_Vollhardt_1989,Jarrell_1992} have emerged as viable alternatives, simplifying many-body problems by approximating complex lattice systems as simpler impurity problems. While this method solves problems in the higher dimensional limit, many interesting results are obtained and comparable to experiments. Particularly, it captures the metal-insulator transition without biased approximation \cite{DMFT_RMP}. The DMFT maps exactly in the infinite spatial dimensions but is approximate for finite dimensions. The approximation can be improved by systematically incorporating corrections from the spatial dependence of the model. There are two major methods to include spatial correlations: perturbative approaches \cite{Toschi_etal_2007,Rubtsov_etal_2009,Fotso2020BeyondQC,Slezak_2009,Hague_etal_2004,Rohringer_etal_2018}, and impurity cluster methods using multiple impurities  \cite{DCA_2000, DCA_Fotso2012, Biroli_Kotliar_2002}.
The latter offers exact calculations up to finite cluster sizes. Recent advancements in numerical algorithms and computing power enable accurate numerical solutions to the single impurity problem. Moreover, in recent years, a novel approach grounded in data science and machine learning has emerged for tackling the single impurity problem\cite{Arsenault_etal_2014,Rigo_etal_2020,Sheridan_etal_2021,Walker_etal_2022,Sturm_etal_2021}. Notably, techniques like dynamical cluster approximation (DCA)  \cite{DCA_2000, DCA_Fotso2012}  and Cellular DMFT (CDMFT) \cite{Biroli_Kotliar_2002} tackle problems involving multiple impurity sites, albeit at the cost of increased computational complexity, representing a significant bottleneck for advancing numerical studies in this field. 

DMFT is often preferred for studying strongly correlated systems like the Anderson Impurity Model(AIM) because it effectively captures local correlations. In systems where local interactions dominate, such as those described by the Anderson impurity model, DMFT provides a powerful framework to treat the local electronic structure accurately while incorporating non-local effects through a self-consistency condition. On the other hand, density matrix embedding theory (DMET), methods such as density functional theory, density matrix renormalization group, and matrix product states partition the system into embedded clusters surrounded by an environment. In addition to addressing local correlations, non-local correlations are accounted for through the embedding environment. This method can capture spatial correlations beyond the single-site level, which may be important in certain systems where the interactions extend beyond neighboring sites. DMFT excels in capturing local correlations and provides a relatively simple framework for studying systems like the AIM. However, DMET offers a more systematic approach to incorporate non-local correlations for systems where such correlations play a significant role \cite{PhysRevB.98.075118}.

Classical computing approaches for accurate solutions of strongly correlated systems face challenges due to exponential scaling, either in computing time (as in the minus sign problem in QMC) or storage (e.g., for exact diagonalization). This limitation hinders predictions for large systems and is unlikely to be overcome by classical hardware or algorithms. However, the emergence of quantum computing offers a new avenue. DMFT, for instance, can be tackled through a hybrid quantum-classical scheme \cite{PhysRevX.6.031045,Keen_etal_2020,kreula2016few,PhysRevResearch.5.023198,jaderberg2022a}, where quantum hardware solves the effective impurity problem while a classical computer post-processes the results. Although direct simulations of models with explicit electron-electron interaction, such as the Hubbard model, near the thermodynamic limit remain beyond current quantum computing capabilities, DMFT simulations on noisy intermediate-scale quantum (NISQ) devices are feasible as the impurity problem can be approximated with a few lattice sites. Nevertheless, computing excited states or Green’s functions remains a challenge for quantum computers in the quantum-classical hybrid scheme ~\cite{Rungger_etal_2019}.

Utilizing quantum computing for solving DMFT poses a unique challenge due to the limited number of available sites or qubits, especially concerning the calculation of the excited state. In the infinite dimension limit, the excitation spectrum becomes a continuous function of energy. But with the constrained Hilbert space dimension, it's represented by discrete delta functions, necessitating a smoothing process for calculations. This issue becomes particularly severe when extracting physical quantities around the Fermi level.

The amalgamation of many-body physics and ML has shown promise in various studies, with classical ML traditionally employed for phase transition detection ~\cite{jaderberg2020minimum,jaderberg2022a,Keen_etal_2020,PhysRevResearch.5.023198, johnston2022perspective,ml_pt}. Quantum machine learning (QML) has emerged as a new paradigm, allowing the representation of input in terms of wavefunctions, fostering the study of phase transitions in many-body systems \cite{wetzel2017unsupervised,van_Nieuwenburg_2017,Casella_etal_2023,Dong_etal_2019,
ml_pt_lat,ml_pt,melting,ml_pom,wetzel2017machine,Walker_Tam_2020,Zhang_etal_2019,Lozano-Gomez_etal_2022,Mano_Ohtsuki_2017,Beach_etal_2019,Ehsan_etal_2020,Baul_2023,Monaco_etal_2023,garcía2022systematic,Wrobel_etal_2022}. Quantum circuits as classifiers, inspired by classical neural networks, have been explored, particularly utilizing convolutional neural networks (CNNs) to identify different phases of matter and their transitions by extracting features from correlation functions of complex systems \cite{keras,tf,Lloyd_etal_2013,Gambs_2008,Baul_2023,Huang_etal_2021,Cappelletti_etal_2020,Belis_etal_2021,Sen_etal_2021,Park_etal_2021,Blank_etal_2020,Schuld_etal_2020,Miyahara_etal_2021,Blance_etal_2021,Grant_etal_2018,LaRose_etal_2020,Du_etal_2021,Abohashima_etal_2020,Chen_etal_2020,Farhi_etal_2018}.

CNNs\cite{Lecun_etal_1998,Beny_2013,Metha_Schwab_2014,Lin_etal_2017,Koch_etal_2020,Funai_Giataganas_2020,Iso_etal_2018,Koch-Janusz_Ringel_2018} have found extensive utility in physics, enabling pattern recognition in statistical models and the analysis of strongly correlated systems ~\cite{ch2017machine,johnston2022perspective}. By incorporating convolutional layers into quantum neural networks, analogous to CNNs, spatial information among qubits can be analyzed effectively, showcasing potential in quantum computing for classifying phases and studying phase transitions ~\cite{ML_RMP2019,Cong_etal_2019}.
%The Quantum Convolutional Layer in QCNN consists of a series of two-qubit unitary operators, which establish correlations between the qubits in the circuit. The Quantum Pooling Layer reduces the number of qubits by performing operations upon each until a specific point and then we disregard some of the qubits in a specific layer. We define a 'pooling layer' where we stop performing operations on certain qubits. Convolution and pooling layers are included in the circuit until the system size (the number of remaining qubits) is reduced significantly. Then, a fully connected layer is utilized as a unitary function on the remaining qubits. The end result is measured by fixing the number of output qubits, and this can be controlled by adjusting the hyperparameters, including the number of convolution and pooling layers.

Quantum-enhanced ML holds great promise for distinguishing different phases of matter shown by recent research\cite{Uvarov_etal_2020,10.21468/SciPostPhys.14.1.005}. The QCNN method might become feasible to recognize various phases of a quantum many-body system, a significant stride toward detecting quantum phase transitions. To apply the QCNN approach, the input needs to be naturally quantum mechanical or generated from some filters that transform the classical data. This eliminates the representation of the wavefunction as a classical vector which grows exponentially with system size\cite{Schindler_etal_2017}. 
The required input should consist of a quantum circuit and the most effective approach for obtaining the wavefunction through the DMFT solution of the two-site single impurity Anderson Model (SIAM) is the variational quantum eigensolver (VQE) method executed on NISQ computers. The main goal of this study is to employ the QCNN to classify the VQE wavefunction obtained via the DMFT for the Hubbard model at the thermodynamic limit. This establishes a framework for detecting quantum phase transition in many-body quantum systems solved with the VQE method.

This paper presents a viable recipe for the study of strongly correlated systems, in particular the quantum phase transition in the thermodynamic limit with the NISQ. Specifically, we focus on drawing the phase diagram. 
The idea combines DMFT, VQE, and QML. The entire process can be implemented on NISQ computers except that the optimization of the parameters is done by classical algorithms. The paper is organized as follows. In Section II, we briefly describe the DMFT approach and its relation to the SIAM. In Section III, the methods and procedure of the two-site DMFT are presented. The quantum simulation of the two-site Hubbard model is described in Section IV. The results for the impurity Green's function, quasi-particle weight, and entanglement entropy are presented in section V 

In section VI, we use the wave function from the DMFT solution to train a QCNN for the detection of the metal-to-insulator phase transition. The results after training the QCNN for the classification of different phases are shown in section VII.
We conclude and propose possible future directions of using VQE and QML methods via DMFT for the study of strongly correlated systems.

\section{Dynamical Mean Field Theory}
Our starting point for strongly correlated electrons is the Hubbard model defined by the Hamiltonian:
\begin{equation}
    H=-\sum_{\langle i,j \rangle \sigma}t_{ij}(\hat{c}^{\dagger}_{i,\sigma}\hat{c}_{j,\sigma}+h.c)+U\sum_{j}\hat{n}_{j,\downarrow}\hat{n}_{j,\uparrow}
    \label{eq:hub}
\end{equation}
where the electrons hop between adjacent lattice sites $i$ and $j$, denoted by $\langle i,j \rangle$, with amplitude $t_{ij} = t$. 
$\hat{c}^{\dagger}_{i,\sigma}$ and $\hat{c}_{i,\sigma}$ respectively denote the creation and annihilation operators for an electron of spin $\sigma$ at site $i$.
$\hat{n}_{j,\sigma}$ is the number of particles with spin $\sigma$ at the site $j$. The interaction between electrons is governed by the on-site Coulomb repulsion, of strength $U$. The detail of the DMFT algorithm is shown in Appendix A.

The key step of the DMFT algorithm is to find Green’s function.
There are two major types of methods to address this
challenge: QMC methods based on sampling the partition function directly or Hamiltonian methods based on
the discretization of the electron bath into a finite number of so-called bath sites. For the latter, the DMFT impurity problem can be described by the Anderson impurity model (AIM) or the single impurity Anderson model
(SIAM).

\begin{figure}
\includegraphics[width=1.00\linewidth]{{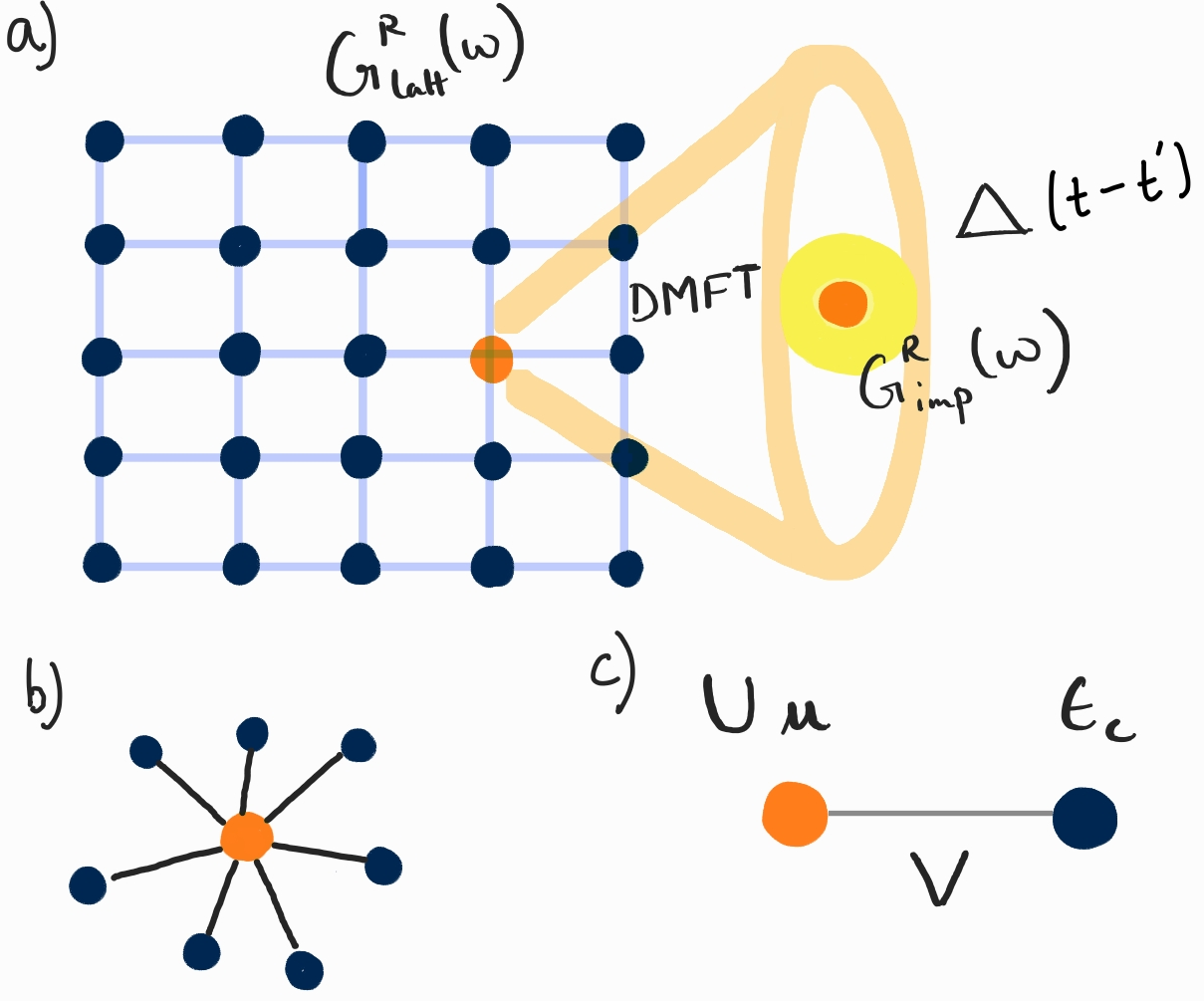}} 
\caption{ (a) DMFT neglects spatial fluctuations, the lattice model with local interaction is mapped into a single site problem and the rest of the lattice is replaced with an effective frequency-dependent mean-field subject to the self-consistency condition. (b) The non-interacting bath sites are connected to the central, interacting impurity site.
(c) Minimal representation of two-site DMFT.
}
\label{dmft}
\end{figure}

The minimal model to mimic the impurity problem is the two-site representation as proposed by Potthoff \cite{PhysRevB.64.165114} as shown in Fig.(\ref{dmft}(c)). It consists of one site representing the impurity and only one site approximately representing the bath.  Refer to Appendix A for the detailed steps.

\section{Methods and Procedure}
After the formulation of the two sites DMFT, the next step is to transform the problem from fermionic operators into spin-${\frac{1}{2}}$ operators for the quantum algorithms. The ground state of the Hamiltonian is obtained by the variational quantum eigensolver (VQE) and the Green's functions are then calculated via the Trotter-Suzuki approximation. With the Green's function, the quasi-particle weight can be calculated and the self-consistency equations can be solved. The procedure is repeated until the desired convergence of the quasi-particle weight is attained. We only consider the half-filled case in this study, so that the chemical potential can be fixed by hand exactly by explicitly enforcing the particle-hole symmetry. We explain the details of the above step in the following. The procedure largely follows that of Kreula {\it et al.} ~\cite{kreula2016few} and modified by Keen {\it et al.} \cite{Keen_etal_2020}. We outline their algorithms here and highlight the differences we propose for calculating the quasi-particle weight.

\subsection{Jordan-Wigner Transformation for Mapping the Fermions to Qubits}
We follow the standard Jordan-Wigner transformation to transform the fermionic creation and annihilation operators into spin operators for representation on a quantum computer. 
A four-qubit system (excluding the ancilla qubit which is used for measurement) is required to represent the two-site SIAM. The spin-down information is encoded by the first two qubits for sites one and two, while the corresponding information for the spin-up occupation is encoded by the third and fourth qubits. This section is provided for the completeness of the paper, the use of Jordan-Wigner transformation to rewrite the Fermionic operators and Green's function for the two-sites model first appeared in Kreula {\it et al.} \cite{kreula2016few}
and later in Keen {\it et al.} \cite{Keen_etal_2020}
The detailed explanation of the mapping can be found in Appendix B.
    \label{eq:G_larger}
%\end{eqnarray}
%\begin{eqnarray}
  %   G^{<}_{imp}(t)&=&\frac{\mathrm{i}}{4} [\langle X_{1}\hat{U}^{\dagger}(t)X_{1}\hat{U}(t) \rangle \nonumber 
 %    +\mathrm{i} \langle X_{1}\hat{U}^{\dagger}(t)Y_{1}\hat{U}(t) \rangle \nonumber \\
%     &-&\mathrm{i} \langle Y_{1}\hat{U}^{\dagger}(t)X_{1}\hat{U}(t) \rangle \nonumber 
%     + \langle Y_{1}\hat{U}^{\dagger}(t)Y_{1}\hat{U}(t) \rangle ] 
         %\label{eq:G_smaller}
%\end{eqnarray}
Following the calculation in Appendix B, we can find the retarded impurity Green’s function $G_{imp}(t)$ can be measured at different times. As the functional form of the two-site problem is known, one can fit the $\mathrm{i}G_{imp}(t)$ on a classical computer. For the two-site DMFT, the interacting Green's function is a four-pole function due to the presence of particle-hole symmetry,
\begin{equation}
    \mathrm{i} G_{imp}(t)=2(\alpha_{1}\cos(\omega_{1}t)+\alpha_{2}\cos(\omega_{2}t)),
  \label{eq:twoSiteG}
\end{equation}
where $\alpha_{2}=0.5-\alpha_{1}$ at half filling. Fourier transforming the above equation leads to
\begin{eqnarray}
  G^{R}_{imp}(\omega+\mathrm{i}\delta)&=&\alpha_{1}(\frac{1}{\omega+\mathrm{i} \delta-\omega_{1}}+\frac{1}{\omega+\mathrm{i} \delta+\omega_{1}}) \nonumber \\ 
  &+& \alpha_{2}(\frac{1}{\omega+\mathrm{i} \delta-\omega_{2}}+\frac{1}{\omega+\mathrm{i} \delta+\omega_{2}}),
  \label{eq:twoSiteG-Omega}
\end{eqnarray}
where $\delta$ is an artificial broadening parameter. As the self-consistency is attained, the Dyson equation is used to calculate the self-energy and, subsequently, the spectral function $A(\omega) =-\frac{1}{\pi}Im[G_{imp}(\omega + \mathrm{i} \delta)]$.

\subsection{Self-energy of the Two-site Model}
A common practice for calculating the self-energy is to employ the Dyson equation in the frequency domain \cite{DMFT_RMP}. Since $G_{0}(\omega)$ is given as the input of the impurity problem and $G_{imp}(\omega)$ is obtained by the above fitting procedure via Fourier transform. The self-energy can be seemingly easy to obtain by 
\begin{equation}
    \Sigma(\omega)=G^{-1}_{0}(\omega)-G^{-1}_{imp}(\omega)
    \label{eq:sigma}
\end{equation}
where the bare Green's function is
\begin{equation}
    G_{0}(\omega+\mathrm{i}\delta)=\frac{1}{(\omega+\mathrm{i}\delta)+\mu-V^{2}/(\omega+\mathrm{i}\delta)}
\end{equation}
Since the quasi-particle weight is given by the derivative of the real part of the self-energy at zero frequency, we need to solve the Eq.(\ref{eq:sigma}) at zero frequency. Unfortunately, it is not easy to calculate the self-energy accurately, as the number of states in the two-sites system is very limited, the subtraction of the two Green's functions in the Dyson equation is essentially a subtraction of a set of delta functions. This does not pose a serious issue when the bath is in the continuum and is actually done routinely in most numerical DMFT calculations \cite{DMFT_RMP}. However, the limited number of states available in the present problem makes the results highly sensitive to the choice of the damping parameter in the Fourier transform of Green's function from the real-time domain to the frequency domain. Strictly speaking, there is also a damping factor associated with time, however, the available time in the present calculation is always limited to a finite number. 

A remedy has been proposed by Keen {\it et al.} in Ref. [\onlinecite{Keen_etal_2020}]. The idea is, instead of calculating the self-energy at zero energy by subtracting the inverse of the bath and the interacting Green's function at zero, to consider the sum over a window of energy. This clearly provides more stable and consistent results which are less dependent on the damping factors~\cite{Keen_etal_2020}. 

This is a viable method for achieving better stability in the iteration step of finding the self-consistent solution of the DMFT. As far as the two-site DMFT is concerned, this integration method is applicable. However, for more general settings such as the cases for multiple bath sites, the self-consistency is more involved. Instead of just the quasi-particle weight, the full self-energy needs to be obtained and the integration method cannot be readily generalized for those situations. Moreover, the integration method also brings an additional parameter that cannot be fixed simply. 

Therefore, it is desirable to have a method that has minimal dependence on arbitrary parameters, specifically the damping and the integration range, but is sufficiently stable for the iterative solution of the DMFT equations. We observe that the system is a finite-size cluster, therefore the self-energy of the two-site impurity problem is a two-pole function of the form \cite{PhysRevB.64.165114},
\begin{equation}
  \Sigma(\omega)=\gamma_{0}+\frac{\gamma_{1}}{\omega-\omega^{s}_{1}}+\frac{\gamma_{2}}{\omega-\omega^{s}_{2}}.  
  \label{eq:Sigma_fit}
\end{equation}

The self-energy is completely determined by five parameters, $\gamma_{0}$, $\gamma_{1}$, $\gamma_{2}$, $\omega^{s}_{1}$ and $\omega^{s}_{2}$, where $\gamma_{1}=\gamma_{2}$ and $\omega^{s}_{1}+\omega^{s}_{2}=0$ at half filling. We can Fourier transform the self-energy back to the time domain with the explicit form of it. The functional form will be the same as that of the Green's function. Instead of calculating the self-energy in the frequency domain, we can use the Dyson equation in the time domain to compute the self-energy. Both the inverse of the bath and interacting Green's function are well-behaved. The parameters, $\gamma_{0}$, $\gamma_{1}$, and $\gamma_{2}$ are then obtained by fitting the self-energy as the difference between the inverse of the bath and the interacting Green's functions in the time domain. After the fitting, the full self-energy is obtained and can be used to extract quasi-particle weight and any other physical quantities. 

The self-energy obtained in this way has a minimal number of arbitrary parameters. The remaining arbitrary parameters are the upper limit of the time and the time step in the Trotter-Suzuki approximation, which are both intrinsic limitations of the quantum algorithm. Therefore, no additional arbitrary parameter is introduced except those limited by the quantum algorithm.

\subsection{Flowchart of the Algorithm}
\begin{figure}
      \includegraphics[width=9.0cm]{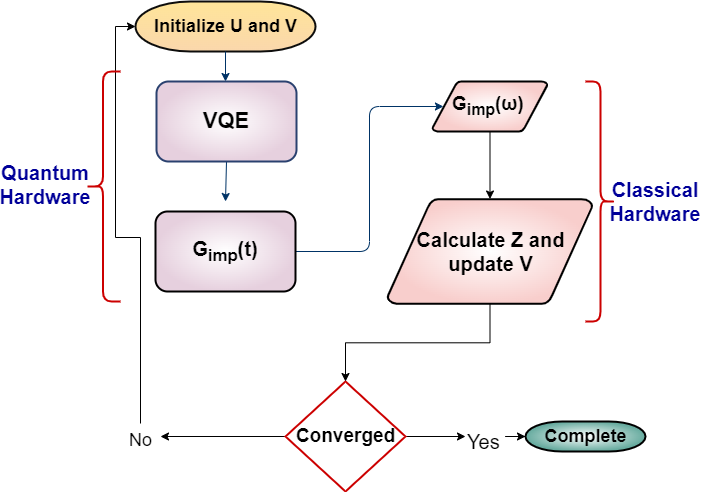}
    \caption{The two-site DMFT calculation flowchart
implemented on a hybrid quantum-classical system. The algorithm was first proposed by Kreula {\it et al.} \cite{kreula2016few} and modified by Keen {\it et al. } \cite{Keen_etal_2020} for the calculation of the quasi-particle weight. 
For the half-filled case, the only external parameter is the Hubbard $U$. The iteration for the self-consistent solution for the quasi-particle weight $\it{\mathcal{Z}}$ starts from a given $U$ and an initial guess for $V$. With these values of $U$ and $V$, the two sites impurity problem is solved by finding the ground state by the variational quantum eigensolver(VQE) and then the Green's function in real-time is obtained by propagating the ground state using the evolution operator $\it{\exp(-\mathrm{i}Ht)}$.
The above procedure can be done in quantum hardware except for the minimization process for the VQE. Once the Green's function in real-time, $G(t)$, is obtained, we find the Fourier transformed Green's function $G(\omega)$ by a fitting procedure as explained in section III-A. With the $G(\omega)$, the quasi-particle weight, $\it{\mathcal{Z}}$, can be calculated by using Eq. \ref{eq:z}. With the obtained $\it{\mathcal{Z}}$, we can update the hybridization $V$, using the relation $V=\sqrt{\it{\mathcal{Z}}}$. Then, the $V$ is checked for convergence. The fitting process is done in the classical hardware. }
\label{fig:flowchart}
  \end{figure}

%The hybrid quantum-classical simulation of the two-site DMFT consists of a few qubit digital quantum simulators to calculate the impurity Green's function and a classical feedback loop where the parameters of the two-site SIAM are updated ~\cite{Keen_etal_2020,kreula2016few}. 
The algorithm, depicted in Fig.(\ref{fig:flowchart}), proceeds as follows~\cite{Keen_etal_2020,kreula2016few}:
\begin{enumerate}
    \item $U$ and $\mu$ are fixed to the desired value in the SIAM and we set the unknown parameters $\epsilon_c=0$ for half-filling, and $V$ to an initial guess.
\item Then we measure the interacting Green's function $\mathrm{i} G_{imp}(t)$ using the technique of single-qubit interferometry.
\item Fourier transform the impurity Green's function by fitting it according to Eq.(\ref{eq:twoSiteG}).
\item 
Calculate the coefficients for the self-energy by using the Dyson equation in the time domain. 
\item  Measure the quasi-particle weight $\it{\mathcal{Z}}$ by using the self energy. 
\item  Update the hopping parameter $V$. 
\item  Repeat steps 2-6 until the self-energy is converged.
\end{enumerate}

\section{Implementation}

\subsection{Ground state Ansatz}
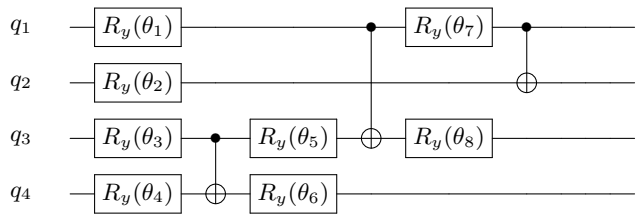
\begin{figure}
\Qcircuit @C=1em @R=.7em {
 q_1 && & \gate{R_{y}(\theta_{1})} & \qw      & \qw                      & \ctrl{2} & \gate{R_{y}(\theta_{7})} & \ctrl{1} & \qw & \qw & \qw & \qw \\
q_2  && & \gate{R_{y}(\theta_{2})} & \qw      & \qw                      & \qw      & \qw                      & \targ    & \qw & \qw & \qw & \qw \\
q_3  && & \gate{R_{y}(\theta_{3})} & \ctrl{1} & \gate{R_{y}(\theta_{5})} & \targ    & \gate{R_{y}(\theta_{8})} & \qw      & \qw & \qw & \qw & \qw \\
q_4  && & \gate{R_{y}(\theta_{4})} & \targ    & \gate{R_{y}(\theta_{6})} & \qw      & \qw                      & \qw      & \qw & \qw & \qw & \qw \\
}\caption{Quantum circuit for the ground state ansatz. We use the same wavefunction for the VQE as in Kreula {\it et al.}\cite{kreula2016few} and Keen {\it et al. } \cite{Keen_etal_2020}.}

\label{fig:VQE}
\end{figure}

The variational ansatz is a quantum circuit with eight single qubit rotations and three CNOT gates, see Fig.(\ref{fig:VQE}). The VQE then optimizes the single qubit rotation parameters to minimize the expectation value of the target Hamiltonian $H_{SIAM}$ for given values of $V, U,\epsilon_c, \mu$ to obtain the ground state of the system.   It's important to understand that there isn't a unique way to choose a variational state. Selecting the optimized variational ansatz is an important but unsolved problem, both in classical variational methods like Variational Monte Carlo (VMC) and in our work. In this study, we're not trying to discover the 'best' wave function~\cite{Keen_etal_2020,kreula2016few}.

The ‘best’ wavefunction varies depending on the model and the parameters involved \cite{Tilly_etal_2022}. Here, we keep the same functional form of the wavefunction across the entire range of on-site interaction strength, even though it may not be the ‘best’ optimized wavefunction. 

\subsection{Retarded Impurity Green's Function}
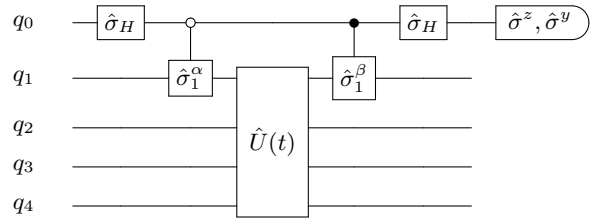
\begin{figure}

\[
\begin{array}{c}
{
\Qcircuit @C=1.0em @R=0.7em {
q_0 && & \gate{\hat{\sigma}_{H}} & \ctrlo{1} & \qw & \ctrl{1} & \gate{\hat{\sigma}_{H}} & \qw & \measureD{ \hat{\sigma}^z , \hat{\sigma}^y }\\
q_1 && & \qw & \gate{\hat{\sigma}^{\alpha}_{1}} &\multigate{3}{\hat{U}(t)} & \gate{\hat{\sigma}^{\beta}_{1}} & \qw & \qw \\
q_2 && & \qw & \qw & \ghost{\hat{U(t)}} & \qw & \qw & \qw  \\
q_3 && & \qw & \qw & \ghost{\hat{U(t)}} & \qw & \qw & \qw  \\
q_4 && & \qw & \qw & \ghost{\hat{U(t)}} & \qw & \qw & \qw 
}
}
\end{array}
\]
\\
\caption{Quantum circuit for measuring the individual components of $G_{imp}(t)$. $\hat{U}(t)$ is composed of quantum gates. $\hat{\sigma}^{H}$ is the Hadamard gate.
$\hat{\sigma_1}^{\alpha}$ and $\hat{\sigma_1}^{\beta}$ can be $X_{1}$ and $Y_{1}$ 
according to the components in Eqs. (\ref{eq:G_larger}) and (\ref{eq:G_smaller}) that we measure. } 
\label{fig:fero}

\end{figure}

The measurement of the impurity Green's function $G_{imp}(t)$ is done by single-qubit Ramsey interferometer, shown in Fig.(~\ref{fig:fero})~\cite{kreula2016few,PhysRevLett.110.230601}, which is used in the more general non-equilibrium case. An ancilla qubit is introduced in addition to the ‘system’
qubits, therefore five qubits are needed to implement the two-site DMFT.

\section{Results}
\begin{figure}
     \centering
     \includegraphics[height=65mm]{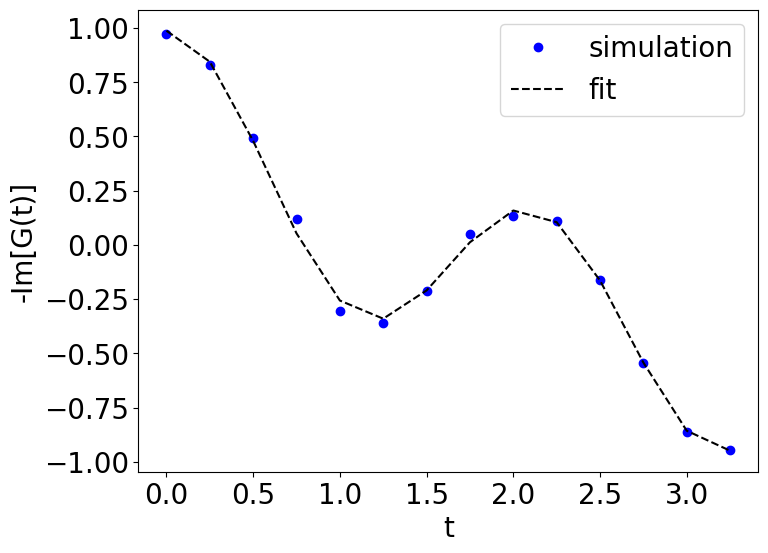}
     \includegraphics[height=65mm]{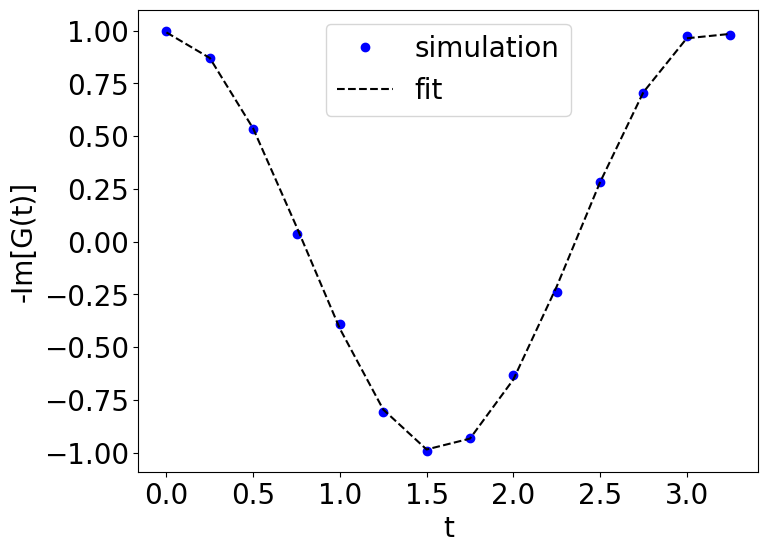}
   \caption{Top: Impurity Green’s function for $U = 8t$ and $V = t$. The parameters for the fit shown are $\alpha_{1}=0.002$, $\alpha_{2}=0.496$, $\omega_{1} = 3.874$, $\omega_{2} = 1.999$. Bottom: Impurity Green’s function for $U = 8t$ and  $V = 0$. The parameters for the fit shown are $\alpha_{1}=0.295$,  $\alpha_{2}=0.055$, $\omega_{1} = 3.143$, and $\omega_{2} = 5.014$. We fit the impurity Green's function to Eq. (\ref{eq:twoSiteG}) 
   } 
   \label{fig:GSigma}
\end{figure}

The results are obtained by a simulator running in classical computers. We start by examining the procedure of obtaining the impurity Green’s function in the DMFT routine. Fig.(\ref{fig:GSigma}) shows the impurity Green’s function for $V = t$ (top) and $V = 0$ (bottom) with $U = 8t$ respectively. We superimpose the data in Fig.(\ref{fig:GSigma}) to the fit obtained according to the exact result of Eq.(\ref{eq:twoSiteG}). The calculated Green's function data is obtained with a modest number of time steps (18 in this case).

\begin{figure}
        \centering
        \includegraphics[height=70mm]{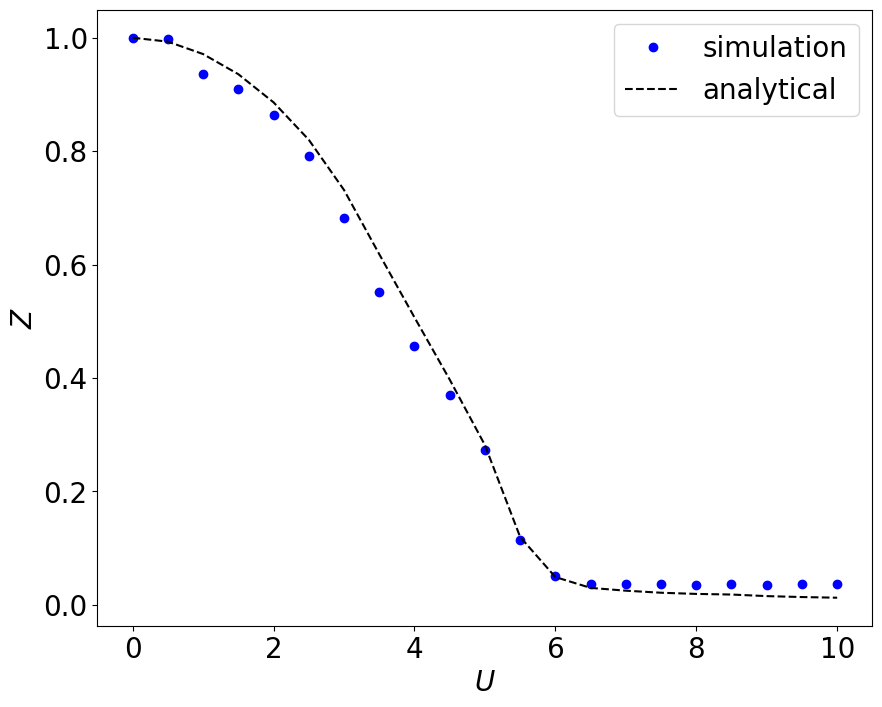}
        \caption{Quasi-particle weight as a function of $U$ at self-consistency from self-energy calculation using Eq.(\ref{eq:z}). The analytical expression of the quasi-particle weight for the half-filling case is obtained from Eq.(31) of Potthoff \cite{PhysRevB.64.165114}}
        \label{fig:Z}
\end{figure}

The impurity Green’s function in the frequency domain $G_{imp}(\omega)$ obtained after self-consistency is achieved, extracted from the fit parameters following Eq.(\ref{eq:twoSiteG-Omega}). %Some of the additional results calculated from the ground state obtained from the VQE are presented in Appendix A. 
We obtain the energies $\omega_{1}$ and $\omega_{2}$ from fitting the Green’s function to Eq.(\ref{eq:twoSiteG}) for half-filling. We then Fourier transform the impurity Green's Function according to Eq.(\ref{eq:twoSiteG-Omega}) to the frequency domain. We use the Dyson equation to calculate the self-energy in the frequency domain from Eq.(\ref{eq:sigma}). We obtain the parameters $\gamma_{0}, \gamma_{1}$, and $\gamma_{2}$, by fitting the self-energy to Eq.(\ref{eq:Sigma_fit}). 
 Fig.(\ref{fig:Z}) shows  the quasi-particle weight at self-consistency using Eq(\ref{eq:z}) for $U$ values ranging from $0.01$ to $10.0$. 

\begin{figure}
        \centering
        \includegraphics[width=90mm]{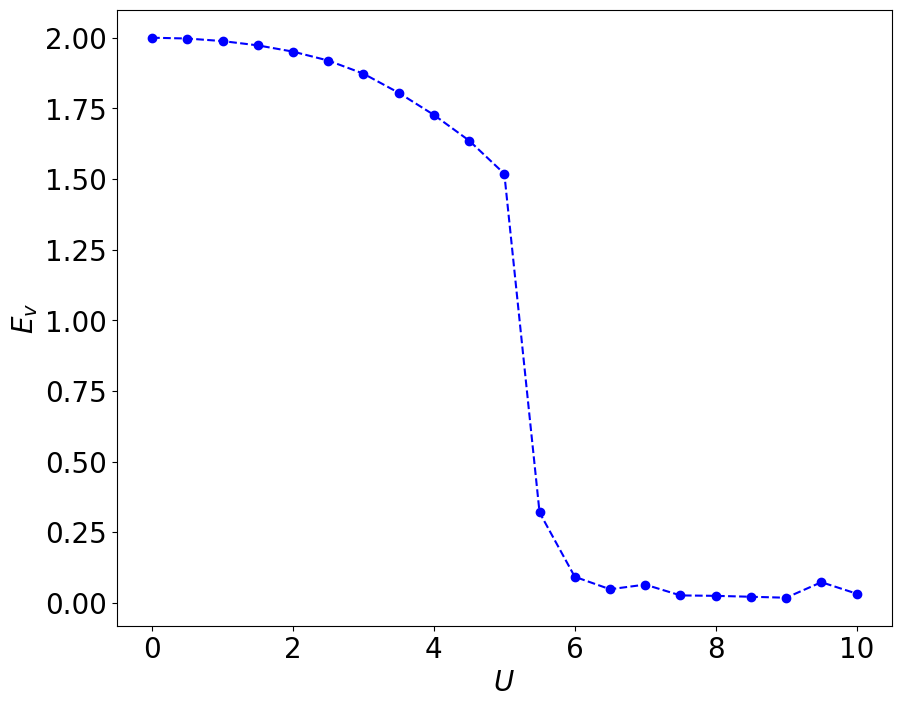}
        \caption{Local entanglement $E_v$ of the two-site SIAM at half-filling versus the on-site coupling $U$ calculated using Eq.(~\ref{eq:entangle}).} 
        \label{fig:Ev}
\end{figure}

In addition to the quasi-particle weight, another quantity that may indicate a phase transition is the entanglement entropy. It has been tested extensively on fermionic lattice models that the entanglement, combined with the finite size scaling, can be used to determine the critical point \cite{Gu_etal_2004,Deng_etal_2006,spalding_etal_2019}. The idea of using entanglement entropy within DMFT has not previously been explored in detail. There is a clear choice of dividing the system into two parts here, as the effective problem that is being solved in DMFT is the SIAM. We can define the entanglement entropy between the impurity site and the bath sites naturally, that is  
\begin{equation}
E_{v} = -Tr[\rho_{imp}\log(\rho_{imp})],
\label{eq:entangle}
\end{equation} 
where $\rho_{imp}$ 
is the reduced density matrix for the impurity site. It is obtained by tracing out the degree of freedom from the bath sites in the density matrix for the ground state ($|GS\rangle$). That is $\rho_{imp} = Tr_{bath}(\rho)$, where $\rho = |GS \rangle \langle GS|$. Please refer to appendix C for the details of the entanglement entropy calculation. As already mentioned, the VQE optimizes the parameters of the quantum circuit shown in  Fig.(\ref{fig:VQE}) to minimize the expectation value of the target Hamiltonian $H_{SIAM}$ for given values of $V, U,\epsilon_c, \mu$ to obtain the ground state $|GS \rangle$ of the system. The behavior of the local entanglement for the half-filling case is shown in Fig.(\ref{fig:Ev}).

In the large $U$ limit, $U \rightarrow \infty $, all sites are singly occupied, one gets $E_{v}(U \rightarrow \infty)=0$. For finite $U$, the hopping process enhances $E_{v}$, and hence reaches its maximum value, $2$ at $U = 0$. There is a sharp change near the value of $U$ corresponds to the Mott transition as estimated from the quasiparticle weight. 


\section{Classification using Quantum Machine Learning}

In the previous section, we obtain the Green's function of the Hubbard model under the DMFT approximation. In addition, we also have the wavefunction for the effective impurity problem from the DMFT, which contains all the information about the system. Instead of directly calculating an ``order parameter", we try to identify the different phases from the wavefunction. It is worthwhile to note that, unlike the DMFT with a continuum bath, the system here is a finite-size system, therefore, there is no true phase transition as that of the thermodynamic limit. Notwithstanding this noteworthy deficit, it has been shown that classical ML methods can locate phase transitions as well as crossovers rather accurately from finite-size classical and even quantum systems \cite{ml_pom,ml_pt}.

\subsection{QML as a Classifier of Wavefunctions}

Since our data is inherently quantum, it is natural to involve a QML approach to try to identify the phase transition. 
In the present study, we focus only on supervised learning. Exactly as in classical ML, the first step is to label input objects \cite{ml_pom}. The input object here is the wavefunction, and the label is whether the wavefunction corresponds to a metal or insulator. 

The major conceptual difference between the QML and the classical ML is in the input data and the function that generates the output label. A popular choice for classical ML is to treat the input data as an array of numbers, and the function is usually chosen as a form of neural network. For supervised learning, a data set consists of input objects, and their corresponding labels are then used to optimize the parameters in the neural network during the training.

Most applications of QML are focused on using classical datasets \cite{Schuld_2021}. We usually convert the wavefunction represented as a classical vector into quantum data. This is the exact process that a quantum classifier requires for classical data, to identify classical images.
For the present study, the input is wavefunctions from the converged DMFT solution, a genuinely quantum dataset, represented in quantum circuits from the VQE.

\subsection{Implentation of the QCNN to Classify the Wavefunctions from the DMFT}

Our objective is to showcase the effectiveness of the QCNN in identifying wavefunctions at different phases. From the converged DMFT solution, we obtain the input wavefunction, and the output label determines whether the wavefunction corresponds to a metal or an insulator.

The quantum neural network ~\cite{Schuld_2021,beer2020training} is a direct conceptual generalization of the classical neural network; the main difference is that the activation functions ~\cite{Schuld_2021,lippmann1994book,agostinelli2014learning} %
are replaced by quantum gates and the input and output are replaced by quantum states instead of an array of classical variables. The quantum neural network we employed in the present study is denoted as QCNN \cite{Cong_etal_2019}. A QCNN consists of two distinct layer types: the pooling layer and the convolution layer. The number of degrees of freedom is reduced by the pooling layer and is substituted by multi-qubit gates, the CNOT gate being the simplest option \cite{Cong_etal_2019}. The convolution layer in the classical CNN is substituted by multi-qubit quantum gates among adjacent qubits. 

The QCNN is built by arranging convolutional and pooling layers in an alternating pattern until the number of pooling layers is reduced to just one qubit or the desired number of qubits. The goal of the present study is to use the QCNN to extract important information from the input quantum data and correctly identify the ``phase" of each wavefunction. We can then use the results to predict the quantum critical point of the Hubbard model via the DMFT. We train the QCNN in a supervised environment, where the correct phase identification for each data point is already known. For a given on-site interaction $U$ at the half-filling, the two-site DMFT approximation can give 
a metallic or Mott-insulator phase. The metallic phase is below $6$ and the Mott-insulator phase is above $6$ \cite{PhysRevB.64.165114}. The reader may refer to the details of the QCNN circuit outlined in Cong {\it et al.} ~\cite{Cong_etal_2019}

\section{Classifier Results using a QCNN}
The QCNN was trained using two different selections of training data. 
200 data points are generated for different $U$ uniformly distributed over $0<U\leq10$. In the first set of training, the QCNN randomly selects $80\%$ of the wavefunctions with labels to indicate their corresponding phases, forming the training data set. In the second set, the data points corresponding to low and high on-site interaction, $U$ were used for training. The metallic phase was assigned the label $-1$, while the Mott insulating phase was assigned the label $+1$. The accuracy of the predictions was benchmarked by bounding the output to $-1$ when the QCNN output measurement was smaller than $0$ and to +$1$ when the measurement was greater than $0$.

\subsection{ Training QCNN with data for randomly picked data for $0.0 < U \leq 10.0$}
\begin{figure}[htpb]
    \centering
        \includegraphics[width=8cm]{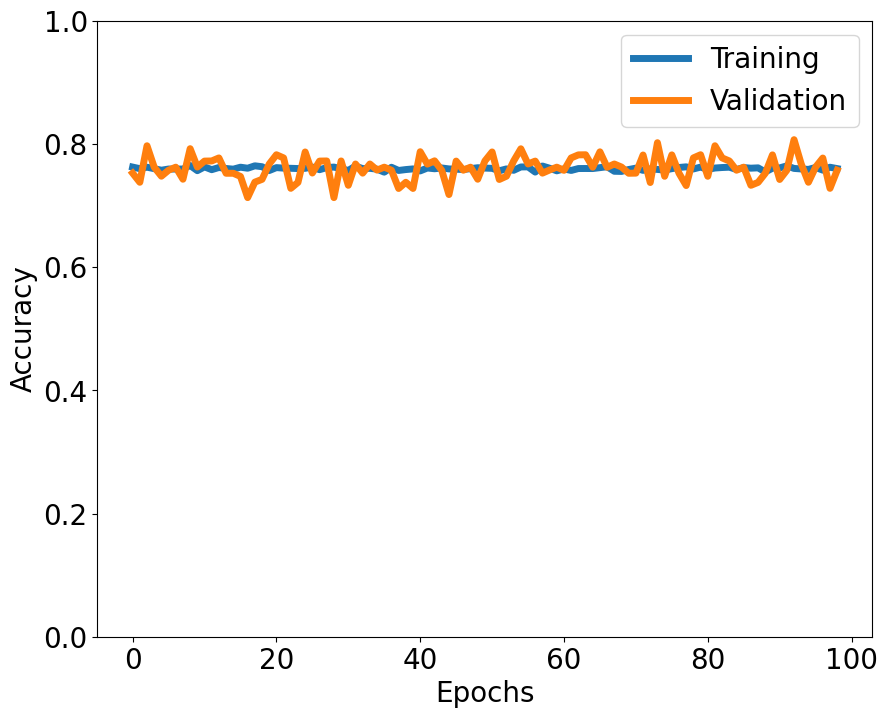}
        \includegraphics[width=8cm]{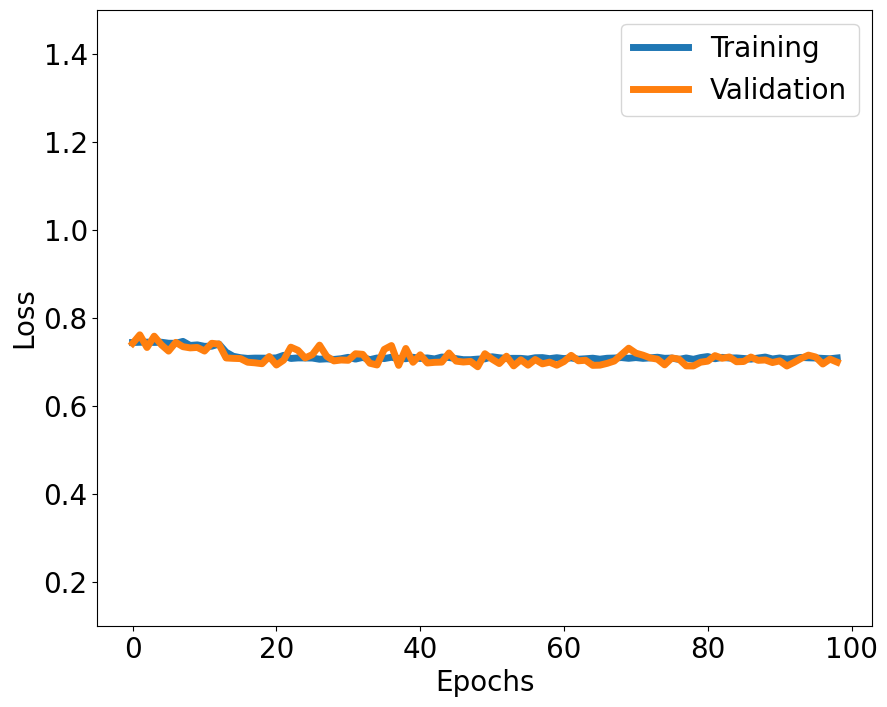}
    \caption{Accuracy and Loss as functions of the number of epochs for both the training and validation datasets in the QCNN. Training with randomly picked data for $0.0 < U \leq 10.0$. Accuracy is the mean of the tensor calculated after checking element-wise equality between the true and predicted labels. Loss is the mean-squared error between the true and predicted labels calculated using Eq.(~\ref{eq:loss}).
    }
    \label{fig:training1}
\end{figure}

We evaluate the performance of our trained QCNN by using the remaining $20\%$ of the samples as a benchmark and track the loss and accuracy at each iteration of the training process. We plot the loss and accuracy as a function of epoch in Fig.(\ref{fig:training1}). The accuracy is calculated by taking the mean of the tensor which is obtained after checking element-wise equality between the true labels and the predictions. We find the loss from calculating the mean of squares of errors between the true labels and the predictions ~\cite{Schuld_2021}.
It is defined as :
\begin{equation}
    L(f(x),y)=\frac{(f(x)-y)^{2}}{N}
    \label{eq:loss}
\end{equation}
where $y$ are the true labels, $x$ are the input feature vectors which are wavefunctions, and $f(x)$ are the output predictions. $N$ is the total number of input samples.

\begin{figure}[h]
    \centering
    \includegraphics[width=8cm]{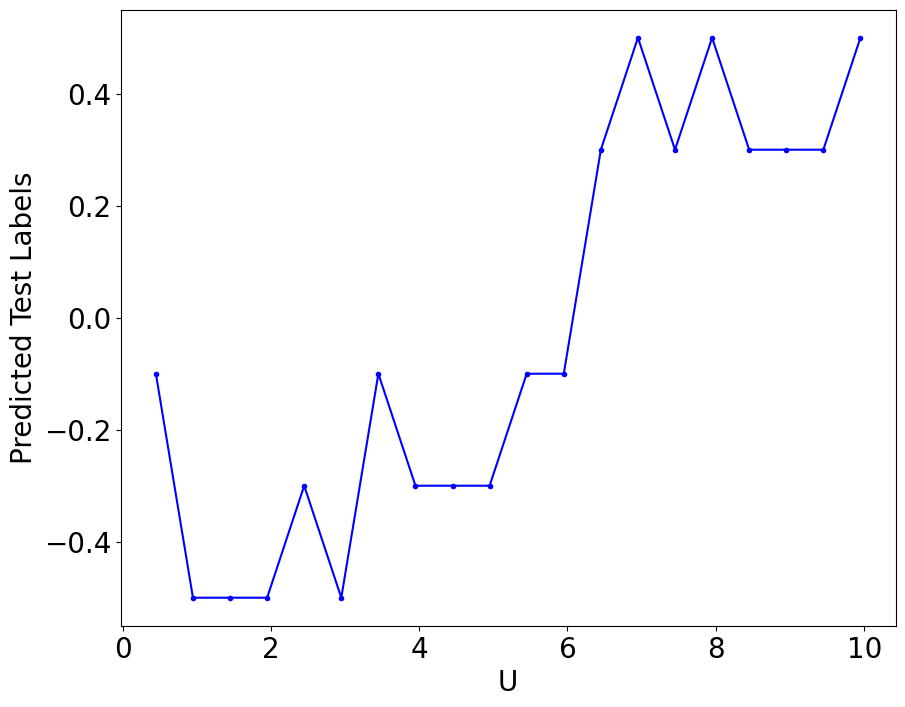}
    \caption{Predicted test labels vs $U$. The predicted labels show a jump from a negative to a positive value around $U=6.0$. Testing randomly picked data for $0.0 < U \leq 10.0$}
    \label{fig:label1}
\end{figure}

By randomly selecting wavefunctions for training, QCNN has the capacity to learn from the complete dataset, encompassing samples from both close and distant regions in relation to the quantum critical point for various $U$ values. This allows the QCNN to become well-versed in the wavefunctions and provide accurate predictions for the trained data. We achieve an accuracy of $75\%$ for both our training and testing datasets, with minimal fluctuations observed throughout the entire process. This demonstrates that randomized training data enables the QCNN to adapt to fluctuations in wavefunctions in the training and generalize well to testing data. 
 The QCNN exhibits the ability to predict the phase of wavefunctions, and we observe the training process through the loss values of the network.  Once the loss converges, indicating no further decrease, the QCNN is considered fully trained and operating at its peak efficiency. Despite the high loss values and insignificant convergence of the loss, each iteration demonstrates the system's training progress. We can visualize the QCNN's training phase and observe its improvement in predicting correctly the phase corresponding to the wavefunctions.  In Fig. (\ref{fig:label1}) we plot the average value of the predicted label as a function of $U$. We average over 10 data points to find the average label, we find that the label passes through zero around $U=6$.
  
\subsection{ Training QCNN with data for $4.0 \geq U$ and $U \geq 7.0$ }
After considering the aforementioned findings, we opted to assess the performance of the QCNN using selected training and testing data. During training, we used wavefunctions that had $U$ values below $4.0$ and above $7.0$, while the test data were within the range of $[4.0-7.0]$. This approach allowed us to determine how effectively the QCNN classifies data close to the known quantum critical point $(U=6.0)$ after training it with data away from it. During training, we obtained an accuracy rate of approximately $75-80\%$, which is not surprising given that the two sets of data with very large and very small values of $U$ are quite different from each other.

\begin{figure}[htpb]
    \centering
        \includegraphics[width=8cm]{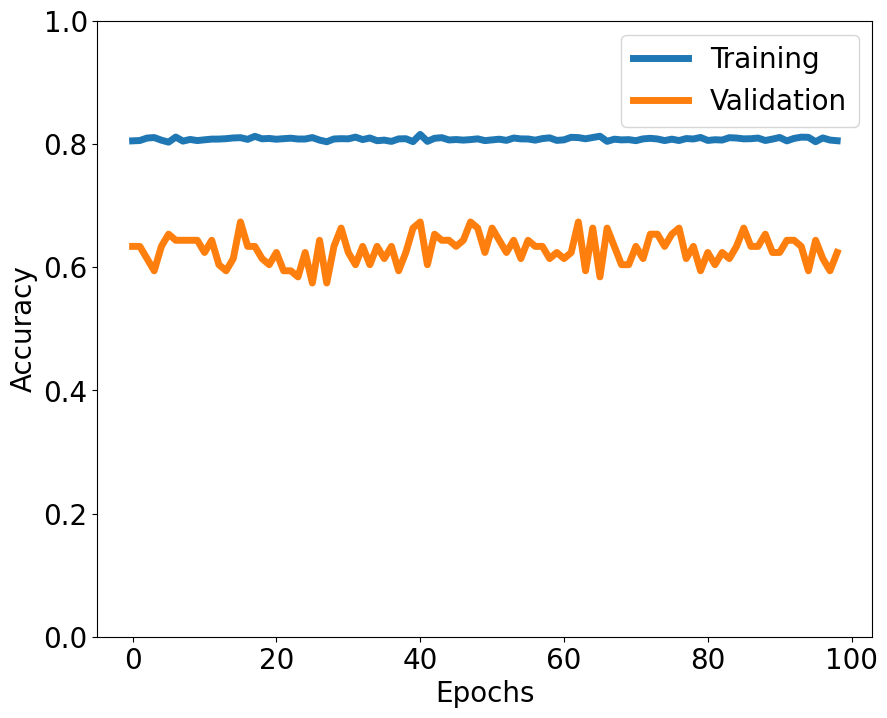}
        \includegraphics[width=8cm]{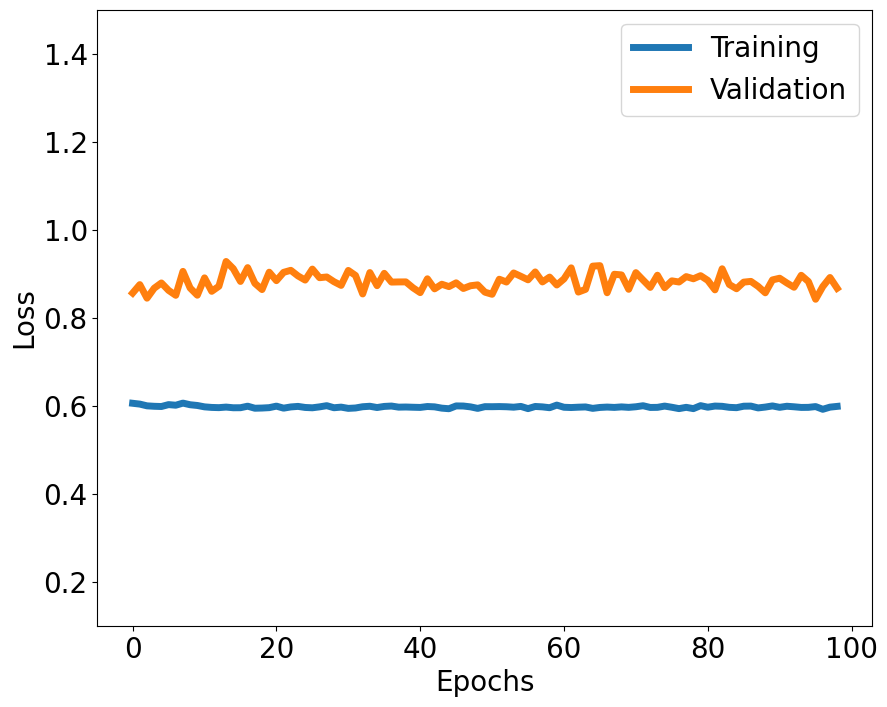}
    \caption{Accuracy and Loss as functions of the number of epochs for both the training and validation datasets in the QCNN. Training with data for $4.0 \geq U$ and $U \geq 7.0$. Accuracy is the mean of the tensor calculated after checking element-wise equality between the true and predicted labels. Loss is the mean-squared error between the true and predicted labels calculated using Eq.(~\ref{eq:loss}). 
    }
    \label{fig:test_acc}
\end{figure}

\begin{figure}[h]
    \centering
    \includegraphics[width=8cm]{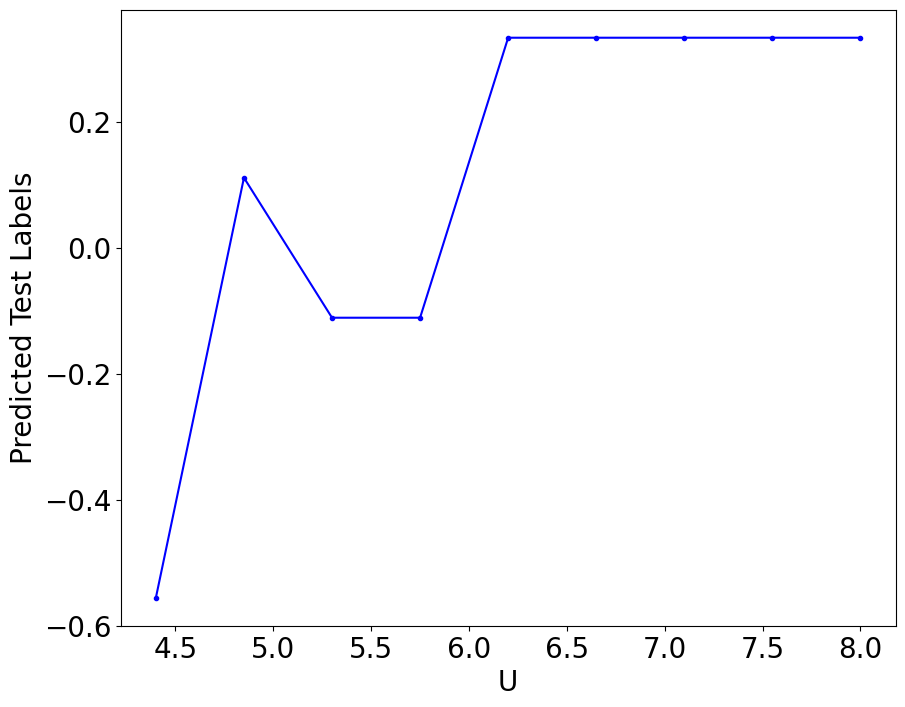}
    \caption{Predicted test labels vs $U$. The predicted labels are hovering around zero for $5.0 \leq U \leq  6.0 $. Training with data for $4.0 \geq U$ and $U \geq 7.0$ and testing data for $4.0 \leq U \leq 7.0$.}
    \label{fig:test_pred}
\end{figure}

We present the results in Fig.(\ref{fig:test_acc}) and Fig.(\ref{fig:test_pred}). We observed that the testing data consistently exhibited a lower accuracy rate than a randomized dataset. The QCNN is somewhat capable of classifying data points that are distant from the quantum critical point. However, as the putative quantum phase transition approached, the predictions became difficult for the network. The average predicted label shows the crossing from negative to positive values for $U$ between $5$ and $6$. The lack of familiarity with data points in this range limited the accuracy of the testing data.  During the training phase, the loss had only slight changes compared to the use of randomized data, and while there was still a decrease, it was minimal. This suggests that the present implementation of the QCNN does not offer significant room for progress when the testing data are isolated from the quantum critical point. 

The exclusive use of high and low values of $U$ during QCNN training has important implications for the identification of quantum critical points. In classical ML, a common technique for detecting phase transitions involves training a supervised model (e.g., a CNN) using control parameters (such as temperature in thermal transitions or external parameters in quantum phase transitions) away from the hypothesized critical point. Our findings indicate that the QCNN can also achieve high accuracy by training on data outside of the hypothesized critical point to predict phases near the critical point \cite{ml_pom}.
Although the predicted label is clearly worse than the case for which the training data span the entire range of $U$. However, the average predicted label suggests that the quantum phase transition occurs within the range of $5 \leq U \leq 6$. Nonetheless, it does not accurately pinpoint the exact value of $U$.  

\section{Discussion}

In this work, we present a hybrid quantum-classical algorithm for calculating the phase transition of the Hubbard model under the DMFT approximation. 
From a practical standpoint, the primary concern is addressing the noise generated by the deep circuit in the Trotter approximation, which affects phase estimation. We did not include the contributions from noise and dissipation in the present study \cite{Rungger_etal_2019}. Recent proposals, in particular, the variational quantum simulation may provide a solution for this problem \cite{Endo_etal_2020,Sakurai_etal_2022}. The major difficulty of the method from a theoretical point of view is how to properly interpret the DMFT approximation realized in a very limited system size for the bath. If the bath is restricted to just one or a handful of sites, it may not be clear whether employing the formulation that presumes a continuous distribution of bath sites is either numerically feasible or physically meaningful. 

The key quantity for the DMFT, particularly in the two-site approximation, is the quasi-particle weight. The quasi-particle weight can be defined as the measure of damping experienced by the quasi-particle at the Fermi level. It is usually calculated using the relation $\it{\mathcal{Z}}=(1- (\partial Re(\Sigma(\omega)) \ /\partial \omega) | \omega =0)^{-1}$.  However, the calculation of self-energy is problematic for a small number of bath sites. This is due to the challenge encountered in solving the Dyson equation for the self-energy $\Sigma(\omega) = G_{0}^{-1}(\omega)-G^{-1}_{imp}(\omega)$. The retarded Green's function is sensitive to the choice of the imaginary part included in the Fourier transform from time to frequency domains. While this may hold true in general, it is often not a significant factor for systems with a continuous bath - i.e., an infinite number of bath sites. In such systems, the choice of imaginary part typically does not play a substantial impact on the results, as they remain consistent across a range of choices.
The focus of the imaginary part is how to distribute the few delta functions for a limited number of sites ~\cite{PhysRevB.64.165114}. The self-energy obtained from the subtraction of the inverse of the full impurity Green's function from the inverse of the bare impurity Green's function cannot be estimated very accurately. 

One method suggested is to replace the calculation of $\mathcal{Z}$ from the Dyson equation at the Fermi energy with an averaging of the integration over an energy window centered around the Fermi level \cite{Keen_etal_2020}. In a way, this approach shifts the problem from selecting the imaginary component in the frequency domain to selecting the integration range over an energy window around the Fermi level. Despite the somewhat arbitrary choice of this range of integration, this method has demonstrated good results \cite{Keen_etal_2020}.

In contrast to many other classical methods used for solving impurity problems, the impurity Green's function obtained through the VQE and Suzuki-Trotter approximation is expressed in real-time and the computation is limited to `zero temperature'. This is an inherent advantage of this hybrid quantum-classical approach compared to conventional Monte Carlo-based approaches. This allows us to estimate the self-energy directly from the difference of the inverse of the interacting Green's function and the bath Green's function without going to the frequency space. The inverse of Green's function is more well-behaved and less prone to error in numerical calculations. 

This approach of calculating the quasi-particle weight directly in the time domain not only avoids the possible ambiguity of choosing the windows of integration but is also more adaptable to a bath with multiple sites. This aligns more broadly with the objective of using hybrid quantum-classical algorithms. The advantage of this approach lies in the fact that we have the information for the self-energy for the full range of the data in the time domain. All the coefficients in the self-energy are calculated and in principle, a full DMFT can be used instead of just the single bath site approximation. In practice, calculations with a large number of bath sites may remain challenging in the near future due to the possibility of increasing errors with larger system sizes. Nevertheless, the formulation presented here can easily be generalized to calculations with many bath sites. 

Besides presenting a modification of the hybrid quantum-classical algorithm for the DMFT, we also tested a new quantity that has not been explored extensively in the context of DMFT \cite{SU_2013}. Conventional DMFT is almost always being studied at a finite temperature, that may be rather low, but almost never at a true zero temperature for the Hubbard model. This is down to the fact that most existing methods for solving the impurity problems are practically feasible at low though non-zero temperatures. As the wavefunction is explicitly calculated, it is technically at the zero temperature and thus we can meaningfully define the entanglement entropy. The use of entanglement entropy as a witness to phase transition has been studied over the past two decades ~\cite{sachdev1999quantum,osborne2002entanglement,osterloh2002scaling}. The general feature is that the entanglement entropy increases as the system is tuned close to a phase transition \cite{Gu_etal_2004}. In the case of the half-filled Hubbard model in the two-site DMFT approach, it is observed that the entanglement entropy reaches its maximum value in the non-interacting limit, but gradually decreases until it saturates once the system enters the Mott insulating phase. A physical interpretation is that in the Mott insulating phase, the bath and the impurity are essentially decoupled, which minimizes the entanglement entropy between them. 

For the present problem, the entanglement entropy may not provide much additional insight. However, there are important problems in condensed matter physics for which there is no obvious order parameter. This has been widely discussed in the context of some heavy fermion materials \cite{Haule_Kotliar_2009}. A method that can detect a putative transition without explicitly constructing an order parameter could be a useful technique.

For capturing the phase transition, we also elaborated on the application of a QML approach to identify phase transitions. The use of ML to locate phase transitions has been well-studied. However, using it for the quantum models requires some manipulations of the data, for example, the wavefunctions from exact diagonalization or the Feynman path integral from QMC are inherent quantum data represented in a classical dataset. Various approaches have been proposed and they are quite successful in finding phase transition \cite{ml_pom}. Since the data from the VQE is true quantum data, it is natural to use QML. The data can be fed into the quantum classifiers without further manipulation of the data. This is an obvious advantage compared to using the classical ML approach for quantum data.  The calculations done here utilizing the wavefunction derived from DMFT's converged solution show promising although imperfect results.The performance stagnation observed in the training process of QCNN is due to the inability to effectively capture the intricate variations present in the data associated with phase transitions.
The selected training data, limited to $U$ values below $4.0$ and above $7.0$, do not provide enough representative samples to effectively train the QCNN to recognize the features associated with phase transitions near the critical point. Hence, the model struggles to generalize accurately to unobserved data within this range.

Since using the ML approach to study phase transitions doesn't require an explicit order parameter, this is advantageous in dealing with crossover problems, where there may not be an explicit order parameter or when the order parameter is unknown. One of them is the crossover between the Fermi liquid and the marginal Fermi liquid \cite{Vidhyadhiraja_etal_2009,Kellar_etal_2023,Yang_etal_2011}. A thorough study of the entanglement entropy may provide more insight into this crossover. Perhaps the major difficulty in studying phase transitions using NISQ computers is the rather limited system size. It's reasonable to imagine that even if one can figure out and calculate an order parameter, a proper finite size scaling as done in conventional numerical simulations may remain out of reach. An ML-based approach could somehow bypass such difficulties and interpret the finite-size results directly. 

\section{Conclusion}
In summary, we provide a complete workflow for using NISQ to study phase transitions of strongly correlated systems in the thermodynamic limit using the dynamical mean field theory (DMFT). We devise a modification of the hybrid quantum-classical method to solve the DMFT using quantum hardware, first proposed by {\it et al.} \cite{kreula2016few} and improved by Keen {\it et al.} \cite{Keen_etal_2020} Our modification provides results without arbitrary choices of parameters in estimating the quasi-particle weight. More importantly, it is feasible to be readily generalized for the DMFT with multiple bath sites. With the ground state wavefunction from DMFT, we proposed two approaches to capture phase transitions. The first one is based on calculating the entanglement entropy between the bath sites and the impurity sites. The second one is based on using the wavefunction from the converged solution of DMFT. Both approaches show promising results in identifying the phase transition under the two-site DMFT approximation.

This method relies on the two-site DMFT approximation, which faces limitations for systems with strong long-range interactions or complex geometries posing computational difficulties in implementing the hybrid quantum-classical method. The framework holds promise for studying phase transitions, but its applicability is more robust in systems with localized interactions and well-defined phases. Future research efforts could focus on addressing these challenges and extending the approach to more diverse classes of strongly correlated materials.

\section{Acknowledgment}
This manuscript is based on work supported by NSF DMR-1728457. This work used high-performance computational resources provided by the Louisiana Optical Network Initiative (http://www.loni.org) and HPC@LSU computing. HFF is supported by the National Science Foundation under Grant No. PHY-2014023.  H.T. is supported by NSF DMR-1944974 grant.  JM is supported by the US Department of Energy, Office of Science, Office of Basic Energy Sciences, under Award Number DE-SC0017861.

%\iffalse

\twocolumngrid

\appendix
\section{Dynamical Mean Field Theory}
DMFT generalizes the mean-field theory for classical systems to the quantum fermionic systems on a lattice. As in conventional mean-field theory, it neglects spatial fluctuations however it explicitly addresses temporal fluctuations.

The action of the Hubbard model on a lattice in $d$ dimension can be written as 
\begin{eqnarray}
&& S= \\ \nonumber -&& \sum_{\boldsymbol{r}_i,\boldsymbol{r}_j,\sigma}\int^{\beta}_{0}\int^{\beta}_{0}d \tau_{i}d \tau_{j} \psi^{\dagger}_{\sigma}(\boldsymbol{r}_i,\tau_i)
G^{-1}_{0}(\boldsymbol{r}_i,\tau_i,\boldsymbol{r}_j,\tau_j) \psi_{\sigma}(\boldsymbol{r}_j,\tau_j) \\  \nonumber +&&U\sum_{\boldsymbol{r}_i}\int^{\beta}_{0}d \tau_{i}\psi{^{\dagger}_{\uparrow}}(\boldsymbol{r}_i,\tau_i)\psi_{\uparrow}(\boldsymbol{r}_i,\tau_i)\psi{^{\dagger}_{\downarrow}}(\boldsymbol{r}_i,\tau_i)\psi_{\downarrow}(\boldsymbol{r}_i,\tau_i)
\label{eq:action}
\end{eqnarray}
where $\psi(\boldsymbol{r}_i,t_{i})$ and $\psi^{\dagger}(\boldsymbol{r}_i,t_{i})$ are Grassmann variables at space-time point $(\boldsymbol{r_i},t_{i})$, $\beta$ is the inverse temperature. The quadratic part of the action corresponds to the kinetic energy, characterized by the bare Green's function $G_0$ which is obtained from the bare dispersion of the Hubbard model. The quartic part corresponds to the interaction from the local coulomb repulsion $U$.

The exact Green's function for this action is characterized by the self-energy $\it{\Sigma}$. In the frequency-momentum space, the Dyson equation gives the relation between the bare Green's function and the exact
Green's function $G$:
\begin{equation}
    G(\boldsymbol{k},\omega)=\frac{1}{G^{-1}_{0}(\boldsymbol{k},\omega)-\Sigma(\boldsymbol{k},\omega)}
    \label{eq:green}
\end{equation} 
 
The full lattice problem, incorporating spatial dependence, is related to a single-site problem through DMFT. The "cavity method" rationalizes this approach in the infinite-dimension limit, where all the degrees of freedom, except for a central site, are integrated out as shown in Fig.(~\ref{dmft}(a)) and Fig.(~\ref{dmft}(b)) \cite{DMFT_RMP}.In the limit of $d \rightarrow \infty$, we rescale the hopping amplitude, $t_{ij} = \frac{t}{\sqrt{2d}}$, ensuring that kinetic energy and interaction energy remain of the same order. The effective action $S_{eff}$ is defined in the frequency domain by:

\begin{multline}
 S_{eff}=-\int d \omega \sum_{\sigma}\psi^{\dagger}_{\sigma}(\omega)\mathcal{G}^{-1}(\omega)\psi_{\sigma}(\omega)\\+U\int_{\omega_1+\omega_3=\omega_2+\omega_4}d \omega_1 d \omega_2 d \omega_3 d \omega_4 \psi^{\dagger}_{\uparrow}(\omega_1)\psi_{\uparrow}(\omega_2) \\
 \psi^{\dagger}_{\downarrow}(\omega_3)\psi_{\downarrow}(\omega_4) . 
\label{eq:imp_action}
\end{multline}
The quadratic part of the impurity action, $\mathcal{G}$ can be written in terms of the lattice Green's function for the Bethe's lattice as \cite{Muller_Hartmann_1989a,Muller_Hartmann_1989b,maier2005quantum,georges1992hubbard,DMFT_RMP}:
\begin{equation}
    \mathcal{G}^{-1}(\mathrm{i} \omega_n)= \mathrm{i} \omega_n+ \mu - t^{2} G(\mathrm{i} \omega_n)
    \label{eq:dmft}
\end{equation}
Eq.(\ref{eq:dmft}) is the self-consistency equation for the DMFT. The Green's function from this action is $\it{G(\mathrm{i} \omega_n)}$, $\omega_n$ are the Matsubara frequencies, which is then fed into the Eq.(\ref{eq:dmft}) until the dynamical mean-field, that is the bath Green's function, $\mathcal{G}(\mathrm{i} \omega_{n})$, is converged ~\cite{DMFT_RMP}.

The key step of the DMFT algorithm is, given this effective action of the system, to find Green's function. There are two major types of methods to address this challenge: QMC methods based on sampling the partition function directly or Hamiltonian methods based on the discretization of the electron bath into a finite number of so-called bath sites. For the latter, the DMFT impurity problem can be described by the Anderson impurity model (AIM) or the single impurity Anderson model (SIAM) defined by the Hamiltonian:

\begin{eqnarray}
H_{AIM}&&=U\hat{n}_{\downarrow}\hat{n}_{\uparrow}-\mu \sum_{\sigma}\hat{n}_{ \sigma} \nonumber \\ 
&&+\sum_{j\sigma}\epsilon_{j}\hat{f}^{\dagger}_{j,\sigma}\hat{f}_{j,\sigma}+\sum_{j\sigma}V_{j}(\hat{c}^{\dagger}_{\sigma} \hat{f}_{j,\sigma}+h.c).
\label{eq:aim}
\end{eqnarray}
Here, $\hat{c}^{\dagger}_{\sigma}$ and $\hat{c}_{\sigma}$ are creation and annihilation operators for impurity electrons, while $\hat{f}^{\dagger}_{j\sigma}$ and $\hat{f}_{j \sigma}$ are those of conduction electrons. The action of the Hamiltonian (\ref{eq:aim}) can be written as that of Eq.(\ref{eq:imp_action}), with $\mathcal{G}$ as the non-interacting Anderson impurity model green's function, if the number of bath electrons goes to infinity. From now on, without specification, we consider single-particle quantities in real-time or frequency. The frequency $\omega$, should be understood as $\omega + \mathrm{i} \delta$.
We can connect the lattice Hubbard model and the single impurity model if we obtain the self-energy as:
\begin{equation}
\int_{\boldsymbol{k}}\Sigma(\boldsymbol{k},\omega)=\Sigma_{imp}(\omega)
\label{eq:sig_equal1}
\end{equation}
This is fulfilled when we reach the DMFT self-consistency condition :
\begin{equation}
 \int_{\boldsymbol{k}} G(\boldsymbol{k},\omega)=G_{imp}(\omega)
\label{eq:sig_equal2}  
\end{equation}

The hybridization and hopping of conduction electrons on a finite 
lattice is related to those of the continuum bath via the bare Green's function as\cite{DMFT_RMP,Muller_Hartmann_1989a,Muller_Hartmann_1989b},   
\begin{equation}
    \mathcal{G}^{-1}(i \omega_{n})= i \omega_{n}+\mu-\Delta(i\omega_{n})
\end{equation}
with
\begin{equation}
    \Delta(i\omega_{n})=\int^{+\infty}_{-\infty}d \omega\frac{1}{(i\omega_{n}-\omega)}\sum_{j \sigma}V^2_{j}\delta(\omega-\epsilon_j).
\end{equation}

This Hamiltonian formulation is more convenient for impurity solvers that are based on matrix diagonalization. Such solvers include exact diagonalization \cite{Caffarel_Krauth_1994,Capone_etal_2007,Liebsch_Ishida_2011}, numerical renormalization group \cite{Bulla_Costi_Pruschke_2008,Peters_etal_2006}, and other more elaborate methods such as density matrix renormalization group, matrix product state\cite{Ganahl_etal_2015,Wolf_etal_2015} and Fock tensor product state \cite{Bauernfeind_etal_2017,Weh_etal_2021}.

\subsection{Two site DMFT}
The minimal model to mimic the impurity problem is the two-site representation as proposed by Potthoff \cite{PhysRevB.64.165114} as shown in Fig.(\ref{dmft}(c)). It consists of one site representing the impurity and only one site approximately representing the bath. The SIAM Hamiltonian for one bath site is as follows.
\begin{eqnarray}
H_{SIAM}=U\hat{n}_{1,\downarrow}\hat{n}_{1,\uparrow}&-&\mu \sum_{\sigma}\hat{n}_{1, \sigma}+\sum_{\sigma}\epsilon_{c}\hat{c}^{\dagger}_{2,\sigma}\hat{c}_{2,\sigma} \nonumber \\
&+&\sum_{\sigma}V(\hat{c}^{\dagger}_{1,\sigma} \hat{c}_{2,\sigma}+h.c),
\end{eqnarray}
$U$ is the on-site Hubbard interaction at the impurity site denoted as site-$1$, and $\mu$ is the chemical potential
for tuning the electrons filling. $\epsilon_c$ and $V$
dare the on-site energy of the non-interacting bath site denoted as site-$2$ and the hybridization between
the impurity and the bath site, respectively. For a mapping of the DMFT to the two-site SIAM, the parameters $\epsilon_c$ and $V$ are determined iteratively so that the two self-consistency conditions below are satisfied. 

In the high-frequency limit, the self-energy of the impurity problem expands in powers of $\it{1/\omega}$ ~\cite{PhysRevB.64.165114}
\begin{equation}
    \Sigma(\omega)=U n_{c}+ \frac{U^{2}n_{c}(1-n_{c})}{\omega} +\mathcal{O}(1/\omega^{2}),
    \label{eq:sigmaHighFreq}
\end{equation}
where $n_{c}=n_{c \sigma}$ is the average occupancy of the impurity orbital:
\begin{equation}
   n_{c}=\langle \hat{c}_{\sigma}^{\dagger}\hat{c}_{\sigma} \rangle =-\frac{1}{\pi}\int^{0}_{-\infty} Im G_{imp}(\omega+i\delta) d\omega.
   \label{eq:siteOccupancyGw}
\end{equation}
Inserting the expansion (\ref{eq:sigmaHighFreq}) into (\ref{eq:siteOccupancyGw}), the $(1/\omega)$ expansion of the on-site lattice Green's function is given as 
\begin{equation}
    G(\omega)=\frac{1}{\omega}+\frac{t_{0}-\mu+Un_{c}^{2}}{\omega^{2}}+\frac{M^{(0)}_{2}+ ...}{\omega^{3}}+\mathcal{O}(1/\omega^{4})
\end{equation}
Here $M^{(0)}_{2}=\Sigma_{j \neq i}t^{2}_{ij}=\int dx x^{2} \rho_{0}(x)$ is the variance of
the bare density of states, $\rho_{0}$. 

Following Potthoff \cite{PhysRevB.64.165114}, the first condition is that the electron filling $n_{imp}=2n_{c}$ of the impurity site is identified with the filling
$n = n_{j \downarrow} + n_{j \uparrow}$ of the lattice model, i.e.,
\begin{equation}
    n_{imp} \equiv n
    \label{eq:n_2sites-dmft}, 
\end{equation}
where we calculate the band filling using
\begin{equation}
   n= -\frac{2}{\pi}\int^{0}_{-\infty} Im G(\omega+i\delta) d\omega
\end{equation}
with $G=G_{imp}$ and a broadening of $\delta$.

In the low-frequency limit, the self-energy of the impurity problem is expanded in powers of $\omega$,
\begin{equation}
    \Sigma(\omega)=a+b \omega +\mathcal{O}(\omega^2)
\end{equation}
with a and b as constants. The quasi-particle weight is
\begin{equation}
    \mathcal{Z}=\frac{1}{1-b}=[1-\frac{\mathrm{d} Re[\Sigma_{imp}(\omega+ i \delta)]}{\mathrm{d} \omega}|_{\omega=0}]^{-1}
    \label{eq:z}
\end{equation}
Neglecting terms of order $\omega^{2}$, yields $G(\omega) = G^{(coh)}(\omega)$ for small $\omega$. $G^{(coh)}(\omega)$ is the coherent segment of the on-site Green's function ~\cite{PhysRevB.64.165114}.  The second self-consistency condition requires comparing the high-frequency expansions of the respective coherent Green's functions \cite{PhysRevB.64.165114,Bulla_Potthoff_2000}.
\begin{equation}
    V^2=\mathcal{Z}M^{(0)}_{2}=\mathcal{Z}\int^{+\infty}_{-\infty}d \epsilon \epsilon^2 \rho_{0}(\epsilon) \equiv \mathcal{Z}t^{\star 2}.
    \label{eq:quasiPart}
\end{equation}

In Eq.(\ref{eq:quasiPart}), $M^{(0)}_2$ denotes the second moment of the bare density of states. The final equivalence is derived from the semicircular density of states of the Bethe lattice ~\cite{bethe1935statistical}.

The primary self-consistency condition holds at lower frequencies up to $\mathcal{O}(\omega^{2})$, by considering the weight, the center, and the variance of the coherent quasi-particle peak. Eq.(\ref{eq:n_2sites-dmft}) and Eq.(\ref{eq:quasiPart}) reformulate the DMFT self-consistency equation. This is sometimes referred to as linear DMFT \cite{Bulla_Potthoff_2000}. Instead of requiring the self-energy as a function of frequency, the linearized DMFT now involves two equations for fixing $\it{\epsilon_{c}}$ and $\it{V}$ as these are the only parameters in the effective two-site model. This is much simplified compared to the full DMFT in which the Green's function for each frequency point from the lattice is mapped to that of the impurity.

\section{Jordan-Wigner Transformation for Mapping the Fermions to Qubits}
We follow the standard Jordan-Wigner transformation to transform the fermionic creation and annihilation operators into spin operators for representation on a quantum computer. 
A four-qubit system (excluding the ancilla qubit which is used for measurement) is required to represent the two-site SIAM. The spin-down information is encoded by the first two qubits for sites one and two, while the corresponding information for the spin-up occupation is encoded by the third and fourth qubits. This section is provided for the completeness of the paper, the use of Jordan-Wigner transformation to rewrite the Fermionic operators and Green's function for the two-sites model first appeared in Kreula {\it et al.} \cite{kreula2016few}
and later in Keen {\it et al.} \cite{Keen_etal_2020}
This process leads to the transformed operators:

\begin{eqnarray}
    c^{\dagger}_{1 \downarrow}&=&\sigma^{-}_{1}=\frac{X_{1}-\mathrm{i} Y_{1}}{2}  \\
    c^{\dagger}_{2 \downarrow}&=&Z_{1}\sigma^{-}_{2}=\frac{Z_{1}(X_{2}-\mathrm{i} Y_{2})}{2}  \\
    c^{\dagger}_{1 \uparrow}&=&Z_{1}Z_{2}\sigma^{-}_{3}=\frac{Z_{1}Z_{2}(X_{3}-\mathrm{i} Y_{3})}{2} \\
    c^{\dagger}_{2 \uparrow}&=&Z_{1}Z_{2}Z_{3}\sigma^{-}_{4}=\frac{Z_{1}Z_{2}Z_{3}(X_{4}-\mathrm{i} Y_{4})}{2}
\end{eqnarray}
    
Here, $X_i, Y_i,$ or $Z_i$ denote operators on the $i^{th}$ qubit, while identity operators act on the remaining qubits without explicitly writing them out. The two-site SIAM Hamiltonian is then given by
\begin{eqnarray}
    H_{SIAM}&=&\frac{U}{4}(Z_{1}Z_{3}-Z_{1}-Z_{3}) \nonumber \\
    &+&\frac{\mu}{2}(Z_{1}+Z_{3})-\frac{\epsilon_c}{2}(Z_{2}+Z_{4}) \nonumber \\
    &+&\frac{V}{2}(X_{1}X_{2}+Y_{1}Y_{2}+X_{3}X_{4}+Y_{3}Y_{4}).
\end{eqnarray}
VQE is employed to find the ground state of the Hamiltonian. Details about the variational wavefunction used are explained in the next section.

Using the first-order Trotter-Suzuki expansion of the time evolution operator acting on the ground state wavefunction to calculate the Green's function in real time. The Trotter-Suzuki transformation is given as  
\begin{eqnarray}
 \hat{U}(t)&=&e^{-\mathrm{i} H_{AIM}t} \nonumber \\
    &\sim& (e^{-\mathrm{i} \frac{V}{2}(X_{1}X_{2}+Y_{1}Y_{2}) \Delta t}e^{-\mathrm{i} \frac{V}{2}(X_{3}X_{4}+Y_{3}Y_{4})\Delta t } \nonumber \\ 
    &\times& e^{-\mathrm{i} \frac{U}{4}(Z_{1}Z_{3})\Delta t} 
   e^{\mathrm{i} \frac{U-2\mu}{4}(Z_{1})\Delta t} e^{\mathrm{i} \frac{U-2\mu}{4}(Z_{3})\Delta t} \nonumber \\
  &\times& e^{\mathrm{i} \frac{\epsilon_c}{2}(Z_{2})\Delta t}e^{\mathrm{i} \frac{\epsilon_c}{2}(Z_{4})\Delta t} + \mathcal{O}((\Delta t)^{2}) )^{N} 
\end{eqnarray}
where $t$ is the total time, $N$ is the number of time steps and $\Delta t = \frac{t}{N}$ ~\cite{PhysRevA.69.010301,PhysRevA.77.066301,PhysRevA.98.032331}.

We only focus on the case $\sigma =\downarrow$ for the first site. As we only consider the paramagnetic solution of the DMFT, it does not break the spin rotational symmetry. The retarded impurity Green's function in the time domain is given as \cite{Freericks_2019}
\begin{eqnarray}
   &\;&  G^{R}_{imp}(t)=\theta(t)(G^{>}_{imp}(t)-G^{<}_{imp}(t))  \\
   &\;&  =-\mathrm{i} \theta(t)(\langle \hat{c}_{1 \sigma}(t)\hat{c}^{\dagger}_{1\sigma}(0)\rangle-\langle\hat{c}^{\dagger}_{1 \sigma}(0)\hat{c}_{1\sigma}(t)\rangle) \nonumber
 \end{eqnarray} 

The above equation is recast into spin operators via the Jordan-Wigner transformation. The Green's functions $G^{>}$ and $G^{<}$ can be written as \cite{kreula2016few,Endo_etal_2020},
\begin{eqnarray}
    G^{>}_{imp}(t)&=&-\frac{\mathrm{i}}{4} [\langle \hat{U}^{\dagger}(t)X_{1}\hat{U}(t)X_{1} \rangle 
    -\mathrm{i} \langle \hat{U}^{\dagger}(t)X_{1}\hat{U}(t)Y_{1} \rangle \nonumber \\
    &+&\mathrm{i}\langle \hat{U}^{\dagger}(t)Y_{1}\hat{U}(t)X_{1} \rangle 
    + \langle \hat{U}^{\dagger}(t)Y_{1}\hat{U}(t)Y_{1}\rangle] 
    \label{eq:G_larger}
\end{eqnarray}
\begin{eqnarray}
     G^{<}_{imp}(t)&=&\frac{\mathrm{i}}{4} [\langle X_{1}\hat{U}^{\dagger}(t)X_{1}\hat{U}(t) \rangle \nonumber 
     +\mathrm{i} \langle X_{1}\hat{U}^{\dagger}(t)Y_{1}\hat{U}(t) \rangle \nonumber \\
     &-&\mathrm{i} \langle Y_{1}\hat{U}^{\dagger}(t)X_{1}\hat{U}(t) \rangle \nonumber 
     + \langle Y_{1}\hat{U}^{\dagger}(t)Y_{1}\hat{U}(t) \rangle ] 
         \label{eq:G_smaller}
\end{eqnarray}

\section{Entanglement Entropy}
We provide the formulas for calculating the single site entanglement entropy. For the spin one-half electron system, each site presents four possible local states: $|0\rangle,|\uparrow\rangle,|\downarrow\rangle,|\uparrow \downarrow\rangle$. The reduced density matrix for the single site %$j$%
can be written as \cite{Gu_etal_2004}
\begin{equation}
    \rho=z|0\rangle \langle 0|+u^{+}|\uparrow\rangle \langle \uparrow|+u^{-}|\downarrow\rangle \langle \downarrow|+w|\uparrow\downarrow \rangle \langle \uparrow \downarrow|.
\end{equation}
with
\begin{eqnarray}
    w&=&\langle 
 \hat{n}_{\downarrow}\hat{n}_{\uparrow} \rangle \\
    &=&\frac{1}{4} \langle 1-Z_{1}-Z_{3}+Z_{1}Z_{3} \rangle \\
    u^{+}&=& \langle \hat{n}_{\uparrow} \rangle-w \\
    &=&\frac{1}{4} \langle 1+Z_{1}-Z_{3}+Z_{1}Z_{3} \rangle \\
    u^{-}&=& \langle \hat{n}_{\downarrow} \rangle -w \\
    &=&\frac{1}{4} \langle 1-Z_{1}+Z_{3}+Z_{1}Z_{3} \rangle \\
    z&=&1-u^{+}-u^{-}-w \\
    &=&\frac{1}{4} \langle 1+Z_{1}+Z_{3}-Z_{1}Z_{3} \rangle.
\end{eqnarray}
With the reduced density matrix, the local von Neumann entanglement entropy can be written as
\begin{equation}
    E_{v}=-z\log_{2}z-u^{+}\log_{2}u^{+}-u^{-}\log_{2}u^{-}-w\log_{2}w
\end{equation}
The local entanglement is determined by the combination of four factors, which collectively influence the system's physical properties. This expression holds greater significance in comprehending the system compared to individual parameters \cite{Gu_etal_2004,Deng_etal_2006,spalding_etal_2019}.

\medskip

%\clearpage

\bibliography{references.bib}

\end{document}